\pdfoutput=1
\documentclass[11pt,twoside,a4paper,cmspaper,final,collab]{cms-tdr}
\def\svnVersion{2e5d915}\def\svnDate{2025/04/24}\def\cmsCernNoTag{CERN-EP-2024-258}\def\cmsCernDate{\today}\def\cmsMessage{Published in the Journal of High Energy Physics as \href{http://dx.doi.org/10.1007/JHEP04(2025)099}{\doi{10.1007/JHEP04(2025)099}.}}
\begin{document}\cmsNoteHeader{TOP-23-005}

\providecommand{\cmsTable}[1]{\resizebox{\textwidth}{!}{#1}}

\newcommand{\PQj}{\ensuremath{\HepParticle{j}{}{}}\xspace}
\newcommand{\PQjpr}{\ensuremath{\HepParticle{j}{}{\prime}}\xspace}
\newcommand{\PQupr}{\ensuremath{\HepParticle{u}{}{\prime}}\xspace}

\newcommand{\pb}{\unit{pb}}
\newcommand{\stt}{\ensuremath{\sigma_\ttbar}\xspace}
\newcommand{\snorm}{\ensuremath{\sigma_\text{norm}}\xspace}
\newcommand{\pp}{\ensuremath{\Pp\Pp}\xspace}
\newcommand{\mt}{\ensuremath{m_{\PQt}}\xspace}
\newcommand{\tW}{\ensuremath{\PQt\PW}\xspace}
\newcommand{\asmz}{\ensuremath{\alpS(m_{\PZ})}\xspace}
\newcommand{\mur}{\ensuremath{\mu_\mathrm{R}}\xspace}
\newcommand{\muf}{\ensuremath{\mu_\mathrm{F}}\xspace}
\newcommand{\abseta}{\ensuremath{\abs{\eta}}\xspace}
\newcommand{\hdamp}{\ensuremath{h_\text{damp}}\xspace}
\newcommand{\Irel}{\ensuremath{I_{\text{rel}}}\xspace}
\newcommand{\wjets}{\ensuremath{\PW\text{+jets}}\xspace}
\newcommand{\deltar}{\ensuremath{\DR}\xspace}
\newcommand{\deltarmed}{\ensuremath{\DR_\mathrm{med}(\PQj,\PQjpr)}\xspace}
\newcommand{\muu}{\ensuremath{m(\PQu,\PQupr)}\xspace}
\newcommand{\ejets}{\ensuremath{\Pe\text{+jets}}\xspace}
\newcommand{\mjets}{\ensuremath{\PGm\text{+jets}}\xspace}
\newcommand{\Njets}{\ensuremath{N_{\text{jets}}}\xspace}
\newcommand{\Nbjets}{\ensuremath{N_{\text{\PQb-jets}}}\xspace}
\newcommand{\njnb}[2]{\ensuremath{#1\PQj#2\PQb}\xspace}

\newcommand{\nSR}{\ensuremath{N^{\mathrm{SR}}}\xspace}
\newcommand{\nCR}{\ensuremath{N^{\mathrm{CR}}}\xspace}
\newcommand{\nSRQCD}{\ensuremath{\nSR_{\mathrm{QCD}}}\xspace}
\newcommand{\nCRobs}{\ensuremath{\nSR_{\mathrm{obs}}}\xspace}
\newcommand{\nCRMC}{\ensuremath{\nSR_{\mathrm{MC}}}\xspace}

\newcommand{\DeepCSV}{DeepCSV\xspace}

\cmsNoteHeader{TOP-23-005}

\title{Measurement of the inclusive \texorpdfstring{\ttbar}{ttbar} cross section in final states with at least one lepton and additional jets with \texorpdfstring{302\pbinv}{302 pbinv} of \texorpdfstring{\pp}{pp} collisions at \texorpdfstring{$\sqrt{s}=5.02\TeV$}{sqrt(s)=5.02 TeV}}

\date{\today}

\abstract{A measurement of the top quark pair (\ttbar) production cross section in proton-proton collisions at a centre-of-mass energy of 5.02\TeV is presented. The data were collected at the LHC in autumn 2017, in dedicated runs with low-energy and low-intensity conditions with respect to the default configuration, and correspond to an integrated luminosity of 302\pbinv. The measurement is performed using events with one electron or muon, and multiple jets, at least one of them being identified as originating from a \PQb quark (\PQb tagged). Events are classified based on the number of all reconstructed jets and of \PQb-tagged jets. Multivariate analysis techniques are used to enhance the separation between the signal and backgrounds. The measured cross section is $62.5\pm1.6\stat\,^{+2.6}_{-2.5}\syst\pm1.2\lum\pb$. A combination with the result in the dilepton channel based on the same data set yields a value of $62.3\pm1.5\stat\pm2.4\syst\pm1.2\lum\pb$, to be compared with the standard model prediction of $69.5\,^{+3.5}_{-3.7}\pb$ at next-to-next-to-leading order in perturbative quantum chromodynamics.}

\hypersetup{%
pdfauthor={CMS Collaboration},
pdftitle={Measurement of the inclusive ttbar cross section in final states with at least one lepton and additional jets with 302 pbinv of pp collisions at sqrt(s)=5.02 TeV},
pdfsubject={CMS},
pdfkeywords={CMS, top quark, top quark cross section}}

\maketitle

\section{Introduction}

The top quark is the heaviest fundamental particle discovered so far. It has a large mass and a large decay width corresponding to a very short lifetime of about $10^{-25}\unit{s}$, a time which is an order of magnitude shorter than the characteristic timescale on which bound state hadrons form. With its unique properties, the top quark potentially carries key information that may help to clarify many open questions in particle physics. The study of top quark pair production (\ttbar) is of crucial importance for precision tests of standard model (SM) predictions and a window to beyond-the-SM physics that may vary the production cross section.

First measurements of the \ttbar production cross section in proton-antiproton collisions were made at the Tevatron at FNAL~\cite{Tevatron}. At the CERN LHC, the ATLAS, CMS, and LHCb Collaborations have performed further measurements in a variety of decay channels at various proton-proton (\pp) collision
energies~\cite{Sirunyan:2017ule,CMS:2021gwv,ATLAS:2022jbj,ATLAS:2022aof,ATLAS:2019hau,Khachatryan:2015uqb,Khachatryan:2016kzg,CMS:2021vhb,Aaij:2018imy,CMS:2023qyl,ATLAS2023slx},
as well as in proton-nucleus~\cite{Sirunyan:2017xku} and nucleus-nucleus~\cite{Sirunyan:2020kov} collisions.
Measurements at various centre-of-mass energies probe different fractions of the proton
longitudinal momentum carried by the gluon~\cite{CMS:2017iqf} and thus can provide complementary information on the parton distribution functions (PDFs), in particular on that of the gluon. In addition, the contributions of the gluon-gluon and quark-antiquark diagram to the production cross section are expected to vary significantly at different centre-of-mass energies, thus making measurements at different $\sqrt{s}$ values an important test of theoretical predictions~\cite{CATANI1996329}.

The first measurement of the \ttbar production cross section (\stt) in \pp collisions at $\sqrt{s}=5.02\TeV$, was performed by the CMS experiment combining the analyses of events with final states containing one or two leptons ($\Pell={}$electron or muon) and at least two jets, using a data sample collected in 2015 that corresponds to an integrated luminosity of 27.4\pbinv~\cite{Sirunyan:2017ule}.
A distinct feature of that data sample is the low number of additional interactions per bunch crossing (pileup) with respect to the standard operating conditions of the LHC. From the experimental side, this implies the important advantage of having a significantly cleaner detector environment, which leads to better detection and reconstruction of the events than in higher pileup scenarios. In 2017, the LHC performed a dedicated run of \pp collisions at $\sqrt{s}=5.02\TeV$ and CMS collected a data sample
corresponding to an integrated luminosity of 302\pbinv. Using these data, CMS obtained an improved measurement of \stt by analyzing dilepton ($\Pepm\PGmmp$) events~\cite{CMS:2021gwv}. The data collected during the same run with the ATLAS detector was used to measure \stt in both the lepton+jets (a single lepton and additional jets) and dilepton final states~\cite{ATLAS:2022jbj}.
This paper presents a new CMS measurement of \stt using events with one lepton and multiple jets and its combination with the measurement in the dilepton final state contained in Ref.~\cite{CMS:2021gwv} ($\stt=60.7\pm 5.8\unit{pb}$), resulting in an improved precision of the measured \stt. The cross section is obtained by performing a maximum likelihood fit to several event categories. The fit is performed by simultaneously adjusting both the cross section and the nuisance parameters associated to the systematic uncertainties.

This paper is organized as follows. A brief description of the CMS detector is given in Section~\ref{sec:CMS}.
Section~\ref{sec:dataandMC} describes the data used for the measurement, as well as the Monte Carlo (MC) simulation samples,
followed by the object reconstruction and event selection in Section~\ref{sec:EventSelection}. The background estimation methods are covered in
Section~\ref{sec:Background}, and the systematic uncertainties in Section~\ref{sec:Systematics}. Results are discussed
in Section~\ref{sec:Results}, and the summary is given in Section~\ref{sec:Summary}.
Tabulated results from this analysis are provided in the HEPData record~\cite{hepdata}.

\section{The CMS detector}
\label{sec:CMS}

The central feature of the CMS apparatus is a superconducting solenoid of 6\unit{m} internal diameter,
providing a magnetic field of 3.8\unit{T}. Within the solenoid volume are a silicon pixel and strip tracker,
a lead tungstate crystal electromagnetic calorimeter (ECAL), and a brass and scintillator hadron calorimeter (HCAL),
each composed of a barrel and two endcap sections. Forward calorimeters extend the pseudorapidity ($\eta$) coverage
provided by the barrel and endcap detectors. Muons are detected in gas-ionization chambers embedded in the steel
flux-return yoke outside the solenoid. A more detailed description of the CMS detector, together with a definition
of the coordinate system used and the relevant kinematic variables, can be found in Refs.~\cite{Chatrchyan:2008zzk,CMS:2023gfb}.

\section{Data and simulated samples}
\label{sec:dataandMC}

The measurement is performed using \pp collision data collected at $\sqrt{s}=5.02\TeV$ with
the CMS detector during a low-energy and low-intensity LHC run in November 2017, corresponding to a total integrated luminosity of 302\pbinv.
The maximum instantaneous luminosity delivered by the LHC during this period was $\mathcal{L}=1.37\times10^{33}\percms$,
and the mean pileup was 2, to be compared with the standard conditions in the 13\TeV run in 2017, where $\mathcal{L}=2.07\times10^{34}\percms$, and the mean pileup was 33.
Events of interest are selected using a two-tiered trigger system~\cite{Khachatryan:2016bia,Sirunyan:2020zal}. The first level (L1),
composed of custom hardware processors, uses information from the calorimeters and muon detectors to select events
at a rate of around 100\unit{kHz} within a fixed latency of 4\unit{\mus}~\cite{Sirunyan:2020zal}. The second level, known as the high-level
trigger, consists of a farm of processors running a version of the full event reconstruction software optimized
for fast processing, and reduces the event rate to around 30\unit{kHz} before data storage. Events considered in this analysis are those in which the trigger detects at least one electron (muon) with transverse momentum $\pt>20$ (17)\GeV.

Signal and background contributions are simulated using several MC generators. These MC simulations are also used to evaluate efficiency and acceptance, and to predict background contributions. The propagation of the generated particles through the CMS detector and the modelling of the detector response in simulations are performed with the
\GEANTfour~\cite{Agostinelli:2002hh} package, using the alignment and calibration obtained from the experimental data.

Simulated \ttbar events are generated at next-to-leading order (NLO) in perturbative quantum chromodynamics (pQCD) using \POWHEG (v2)~\cite{Frixione:2007vw,Alioli:2010xd,Frixione:2007nw,Nason_2004}, assuming a top quark mass (\mt) of 172.5\GeV.

The NNPDF3.1~\cite{Ball:2017nwa} next-to-next-to-leading order (NNLO) PDFs are used.
The simulation of the single top quark production in the $t$ channel and in association with a \PW boson (\tW) is also performed with \POWHEG, using the same setup.

{\tolerance=800
The \MGvATNLO (v2.4.2) generator~\cite{Alwall:2014hca},
is used to simulate Drell--Yan (DY) quark-antiquark annihilation into lepton-antilepton pairs through \PZ boson or virtual photon exchange, and \PW boson production with additional jets (\wjets). The simulation is performed at NLO in pQCD and includes up to two extra partons at the matrix element (ME) level. The FxFx matching scheme~\cite{Frederix:2012ps} is used to match jets from the ME calculations and the parton showering (PS). For all the processes, simulated events are interfaced with {\PYTHIA8} (v230)~\cite{Sjostrand:2014zea}
with the ``CP5'' tune~\cite{Sirunyan:2019dfx} for PS, hadronization, and the
underlying event description.
\par}

The SM prediction for \stt is calculated with the \TOPpp program~\cite{Czakon:2011xx} at NNLO in pQCD including soft-gluon resummation at next-to-next-to-leading logarithmic (NNLL) accuracy~\cite{Czakon:2013goa} for $\mt=172.5\GeV$ and a strong coupling at the \PZ boson mass, \asmz, of $0.118\pm0.001$~\cite{PDG2020}. At 5.02\TeV, this prediction is $69.5^{+2.0}_{-2.3}\,\text{(scale)} \pm 2.9\,\text{(PDF+\alpS)}\pb$.
The first uncertainty reflects variations in the factorization (\muf) and renormalization (\mur) scales.
The second uncertainty is associated with possible choices of the PDFs and the \alpS value, calculated using the PDF4LHC prescription~\cite{Butterworth:2015oua}, that encompasses the MSTW2008 68\% CL NNLO~\cite{martin2009parton,Martin_2009}, CT10 NNLO~\cite{Lai_2010,Gao_2014}, and NNPDF2.3 5f FFN~\cite{Ball_2013} PDF sets. The cross section predictions used for the normalization of $t$ channel and \tW events are calculated at NLO in both QCD and electroweak (EW) orders of approximation matched with PS resummation~\cite{Campbell:2020fhf, Kidonakis:2021vob}, whereas for \wjets and DY events the calculation is performed at NNLO in QCD and NLO in EW~\cite{Melnikov:2006kv}. The signal and background MC samples used in the analysis are summarized in Table~\ref{tab:mc}, as well as the cross section predictions used for the normalization of events.

\begin{table}[!ht]
\centering
\topcaption{%
Summary of MC samples used to model the signal and background processes. The last column corresponds to the pQCD and EW order of approximation used to normalize the distributions provided by the
generators. The predictions for \ttbar and $t$ channel and \tW single top quark production are calculated with the PDF4LHC prescription~\cite{Butterworth:2015oua}. Approximate NNLO refers to NLO calculation matched with PS resummation.}
\label{tab:mc}
\renewcommand{\arraystretch}{1.3}
\cmsTable{\begin{tabular}{cccc}
    \multirow{2}{*}{Process} & \multirow{2}{*}{Generator + Parton Shower} & Cross section, & \multirow{2}{*}{Order of \snorm approximation} \\[-5pt]
    & & \snorm (pb) & \\
    \hline
    \ttbar & \POWHEG + \PYTHIA8 & $69.5\pm3.6$ & NNLO + NNLL~\cite{Czakon:2011xx,Czakon:2013goa} \\
    $t$ channel & \POWHEG + \PYTHIA8 & $30.3\pm0.6$ & Approximate NNLO~\cite{Campbell:2020fhf} \\
    \tW & \POWHEG + \PYTHIA8 & $6.54\pm0.36$ & Approximate NNLO~\cite{Kidonakis:2021vob} \\
    \wjets & \MGvATNLO + \PYTHIA8 & $21\,160\pm330$ & NNLO [QCD] + NLO [EW]~\cite{Melnikov:2006kv} \\
    DY & \MGvATNLO + \PYTHIA8 & $3\,647\pm63$ & NNLO [QCD] + NLO [EW]~\cite{Melnikov:2006kv} \\
\end{tabular}}
\end{table}

\section{Object reconstruction and event selection}
\label{sec:EventSelection}

The particle-flow (PF) algorithm~\cite{CMS-PRF-14-001} aims to reconstruct and identify each individual particle in an event, with an optimized combination of information from the various elements of the CMS detector. The energy of photons is obtained from the ECAL measurement. The energy of electrons is determined from a combination of the electron momentum at the primary interaction vertex (PV) as determined by the tracker, the energy of the corresponding ECAL cluster, and the energy sum of all bremsstrahlung photons spatially compatible with originating from the electron track. The energy of muons is obtained from the curvature of the corresponding track. The energy of charged hadrons is determined from a combination of their momentum measured in the tracker and the matching ECAL and HCAL energy deposits, corrected for the response function of the calorimeters to hadronic showers. Finally, the energy of neutral hadrons is obtained from the corresponding corrected ECAL and HCAL energies.

Jets are clustered from all objects reconstructed by the PF algorithm using the anti-\kt algorithm ~\cite{Cacciari:2008gp, Cacciari:2011ma} and a distance parameter of 0.4 is used. The jet momentum is determined as the vectorial sum of all particle momenta in the jet, and is found from simulation to be, on average, within 5--10\% of the true momentum over the whole \pt spectrum and detector acceptance. Jet energy corrections are derived from simulation studies so that the average measured energy of jets becomes identical to that of particle-level jets. Particle-level jets are the result of adding PS and hadronization information to particles produced after the generation of the hard-scattering process. To determine any residual differences between the jet energy scale in data and in simulation, measurements of the momentum balance are used, leading to appropriate correction factors~\cite{Khachatryan:2016kdb}. Additional selection criteria are applied to each jet to remove jets potentially dominated by instrumental effects or reconstruction failures~\cite{CMS:2017wyc}. Jets are identified as coming from the fragmentation of bottom quarks using the medium working point of the \DeepCSV \PQb tagging algorithm~\cite{Sirunyan:2017ezt}. With this working point, we obtain for \ttbar events a \PQb tagging efficiency of 68\% for \PQb jets, and mistag rates of 12 and 1.1\% for \PQc and light jets with $\pt>20\GeV$. Throughout the full text, we will be referring to jets in general as '\PQj', and to non-\PQb-tagged jets as '\PQu'.

A multivariate discriminator based on the shower shape and track quality of the electron candidates~\cite{CMS:2020uim} is used to identify electrons. Electron candidates are required to satisfy $\abseta<2.4$ (except for the region $1.479<\abseta<1.566$, which is excluded because that is the transition region between barrel and endcaps, where the electron reconstruction is suboptimal) and $\pt>20\GeV$. A veto is applied on candidates that are matched to a secondary vertex consistent with a photon conversion, or have a missing hit in the inner layer of the tracker. Reconstructed muon candidates are required to have $\abseta<2.4$ and $\pt>20\GeV$, and must fulfil criteria on the
geometrical matching between the tracks reconstructed by the silicon tracker and the muon system, and on the quality of the global fit~\cite{Sirunyan:2018fpa}. Both electron and muon energy scales and resolution are corrected according to measurements using \PZ boson events in simulation and low-pileup \pp collision data at 5.02\TeV~\cite{CMS:2020uim,Sirunyan:2018fpa}.

Pileup collisions may cause events to contain multiple interaction vertices. The PV is taken to be the vertex corresponding to the hardest scattering in the event, which is evaluated using tracking information alone, as described in Section~9.4.1 of Ref.~\cite{CMS-TDR-15-02}.

Lepton candidates must be consistent with originating from the PV. This is guaranteed by requiring the longitudinal impact parameter to be smaller or equal to 0.10\unit{cm} and the transverse impact parameter smaller than 0.02 (0.05)\unit{cm} for electrons (muons).
Electrons and muons must also satisfy
a requirement on their relative isolation (\Irel), defined as the scalar \pt sum of all the particles inside a cone around the lepton direction, excluding the lepton itself, divided by the lepton \pt.
The cone size, defined as $\DR=\sqrt{\smash[b]{(\Delta\eta)^2+(\Delta\phi)^2}}$, where $\phi$ is the azimuthal angle, changes as a function of the lepton \pt as $\DR(\pt)=10\GeV/\pt$ if $50<\pt<200\GeV$, $\DR=0.2$ if $\pt<50\GeV$, and $\DR=0.05$ if $\pt>200\GeV$.
Finally, electrons (muons) must satisfy the condition on isolation $\Irel<0.085$ (0.325).

A gradient boosted decision tree (BDT) is used to distinguish leptons coming from hadron decays and misidentified leptons, collectively denoted ''nonprompt'' leptons, from those produced in the decays of weak bosons (''prompt''). The BDT is trained using MC simulations and exploits the properties of the jet which contains the lepton, as well as properties of the lepton itself, as input variables~\cite{muonMVA}.
Leptons associated with a jet satisfying the loose working point of the \DeepCSV \PQb tagging algorithm~\cite{Sirunyan:2017ezt} are rejected to further suppress nonprompt leptons originating from \PQb quark decays. Using this working point with \ttbar events, we obtain a \PQb tagging efficiency of 84\%, and mistag rates of 41 and 11\% for \PQc and light jets with $\pt>20\GeV$.

The \ttbar candidate events are required to contain a single selected electron or muon, and at least 3 jets. A veto on any second lepton of opposite flavour with respect to the selected one, with \pt as low as 10\GeV, is applied to ensure the orthogonality to the dilepton analysis~\cite{CMS:2021gwv}.
Only jets with $\pt>25\GeV$, $\abseta<2.4$, and containing no selected leptons are considered. Furthermore, events are required to ensure that the magnitude of the missing transverse momentum \ptvecmiss, defined as the negative vector \pt sum of all identified physics objects~\cite{ptmissperformance}, is greater than 30\GeV. This is done in order to reduce the background contamination from SM events composed uniquely of jets produced through the strong interaction, referred to as QCD multijets events. Events passing these selection requirements are classified into eight categories depending on the number of jets (\Njets) and \PQb-tagged jets (\Nbjets) and on the flavour of the lepton (electron or muon). Separate categories are defined for events with exactly 3 and $\geq$4 jets, which are further divided into categories of one \PQb-tagged jet and $\geq$2 \PQb-tagged jets. Throughout this document the jet multiplicity categories will be denoted as \njnb31, \njnb41, \njnb32, and \njnb42.

\section{Background estimation}
\label{sec:Background}

The event selection is designed to maximize the signal contribution over the expected background contamination. The background contributions come from the following processes: single top quark production in association with a \PW boson, \wjets production in the single-lepton final state, and QCD multijet events when a jet is misidentified as a lepton. In addition, DY events in three cases: when one of the two leptons in the final state is misidentified as a jet, if there is a lepton outside of the detector acceptance, or if a \PZ boson decays into two \PQt leptons and there is a single leptonic \PQt decay.

The estimation of these background contributions is based on MC simulation, except for the QCD multijet contribution, which is determined from control samples in data.

\subsection{QCD background estimate}

The QCD multijet background contribution is estimated from data using events with non-isolated leptons that are not selected by the lepton BDT described in Section~\ref{sec:EventSelection}.
The evaluation of the QCD background follows the method used in Ref.~\cite{Sirunyan:2017ule}. A control region (CR) with events containing non-isolated leptons is defined by inverting both the isolation (\Irel) and the lepton BDT requirements. To estimate the number of QCD background events (\nSRQCD) in the signal region (SR), an extrapolation factor is calculated and applied to the QCD yield in the CR. To do so, the following formula is used:
\begin{equation}\label{eq:qcdextrapolation}
  \nSRQCD=\big(\nCRobs-\nCRMC\big)\frac{\nSR_{\text{obs, low \ptmiss}}-\nSR_{\text{MC, low \ptmiss}}}{\nCR_{\text{obs, low \ptmiss}}-\nCR_{\text{MC, low \ptmiss}}}.
\end{equation}
QCD multijet events in the CR are determined as the number of events in data (\nCRobs) subtracted by the contribution, derived from the MC samples, of the signal and the other background sources considered (\nCRMC). The factor is determined using events with $\ptmiss<20\GeV$, where a large contribution of QCD background and low signal contamination is expected, as signal events contain genuine  \ptvecmiss coming from a high-\pt neutrino.

To evaluate the normalization uncertainty in the QCD background contribution, the normalization factor is determined
after varying the \ptmiss requirement by $\pm$5\GeV in both the SR and CR.
A flat normalization uncertainty of 30\% is applied to the QCD contribution across all event categories since it covers most of the variations. It is worth noting that the QCD background contribution is quite small in this analysis, as can be seen in the distributions shown in Figs.~\ref{fig:inputsdatamc}--\ref{fig:prefitmu}.

\subsection{\texorpdfstring{\wjets}{W+jets} background discrimination}

The final state with four quarks from the top quark decays leads to the expectation of at least four jets in signal events. Due to acceptance and selection efficiency effects, however, a relevant number of signal events has only three selected jets, thus contributing to the \njnb31 category. In this category, a larger contribution from \wjets events is expected when compared to others. Since jets in this process come mainly from radiation, for higher jet multiplicity categories they are softer and more forward, thus not passing the selection criteria. Therefore, a multivariate-analysis (MVA) classifier based on random forests~\cite{randomforest} is trained to separate the \ttbar signal from the \wjets background. The classifier is trained using subsets of the \ttbar and \wjets simulated samples, orthogonal to those used for the analysis. In these subsets, 70\% of events are used for the training and the remaining events for testing. The training is performed inclusively in the $\Pell$+3j1b categories. After an optimization procedure, the hyperparameters are fixed to 500 trees with a maximum depth of 6. The classifier is trained using the variables described in Table~\ref{tab:variablesmva}, where they are ordered according to their discriminating power.

\begin{table}[!ht]
\centering
\topcaption{%
    Summary of the variables used for the training of the MVA classifier in the \njnb31 categories. The variables are ordered according to their discriminating power, assessed via the mean decrease in impurity method~\cite{cutler2012random}. `Leading' jet refers to the jet with the highest \pt in the event.
}
\label{tab:variablesmva}
\begin{tabular}{cl}
    Variable & Definition \\
    \hline
    \deltarmed & Median \deltar between all possible combinations of two jets  \\
    \muu & Invariant mass of the two non-\PQb-tagged jets  \\
    $\DR(\PQu,\PQupr)$ & \deltar between the two non-\PQb-tagged jets  \\
    $m_\mathrm{min}(\PQj,\PQjpr)$ & Minimum invariant mass of all possible combinations of two jets  \\
    $m(\Pell,\PQb)$ & Invariant mass of the lepton and the \PQb-tagged jet  \\
    \HT & Scalar \pt sum of all jets in the event  \\
    $\DR(\Pell,\PQb)$ & \deltar between the lepton and the \PQb-tagged jet  \\
    $\pt(\PQj_0)$ & \pt of the leading jet  \\
\end{tabular}
\end{table}

\begin{figure}[!ht]
\centering
\includegraphics[width=0.48\textwidth]{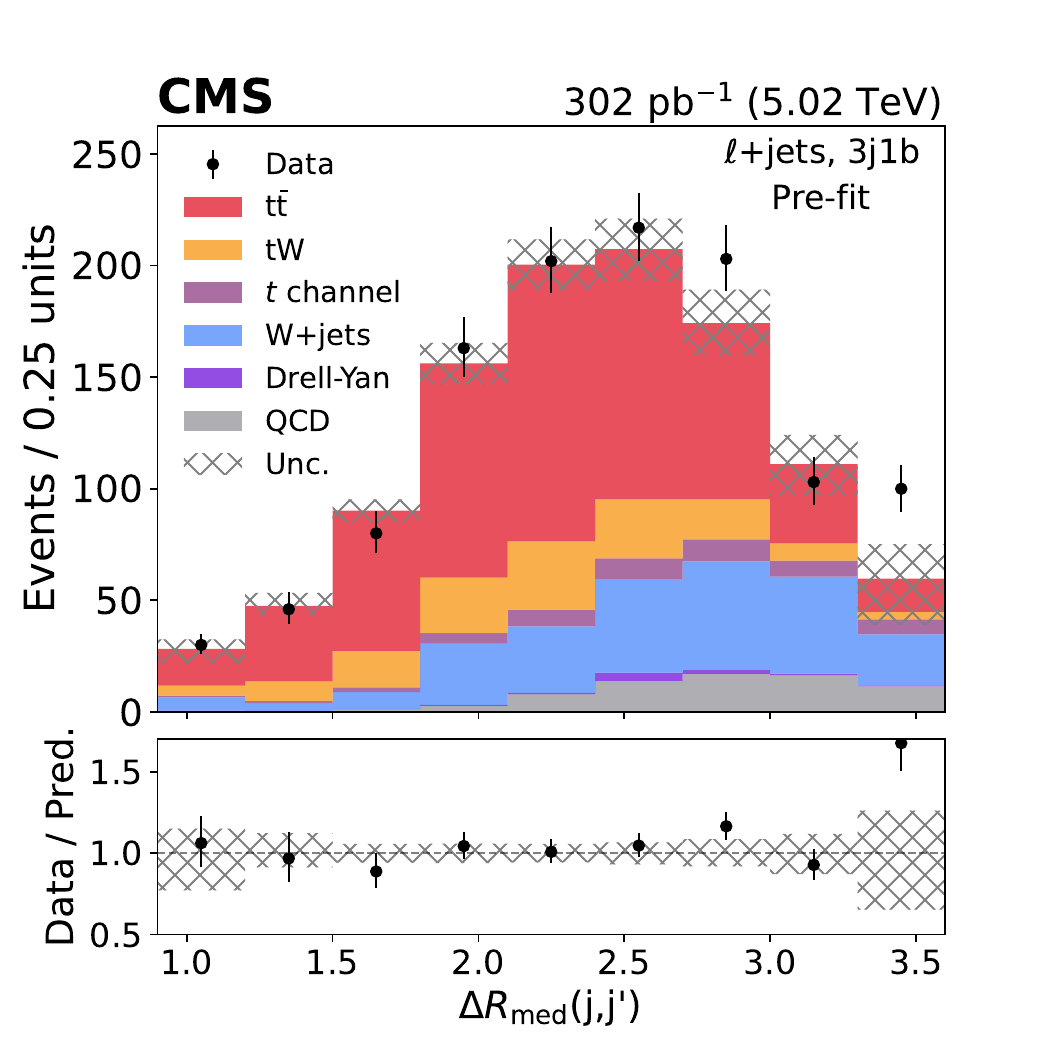}
\hfill
\includegraphics[width=0.48\textwidth]{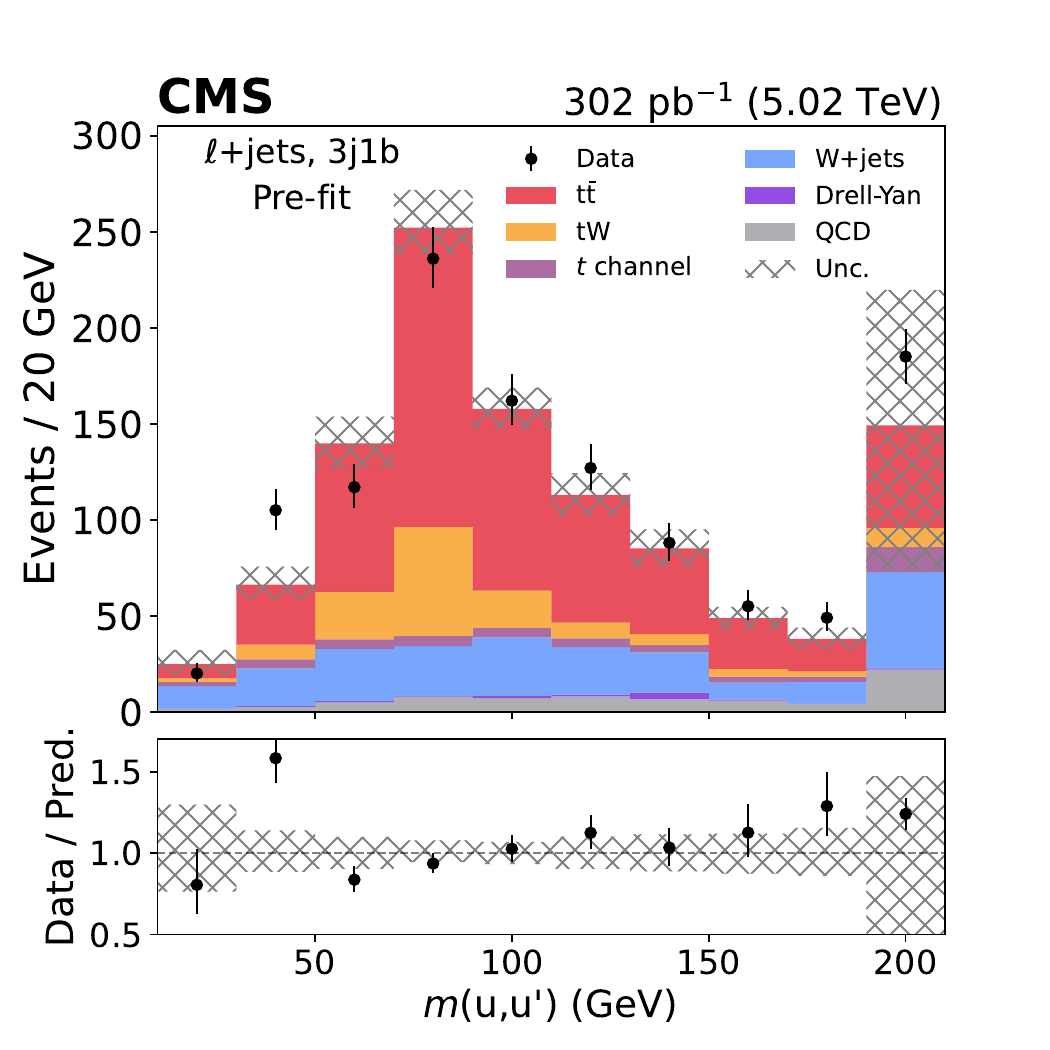}
\caption{Distributions for data and expected signal and background contributions of the most discriminating input variables used for the random forest training, \deltarmed (left) and \muu(right), in the \njnb31 category, before the maximum likelihood fit. The vertical error bars represent the statistical uncertainty in the data, and the shaded band the uncertainty in the prediction. All uncertainties considered in the analysis are included in the uncertainty band. The lower panels show the data-to-prediction ratio. The first and last bins in each distribution include underflow and overflow events, respectively. The normalizations are taken with respect to the SM predictions except for QCD, which is estimated from data.}
\label{fig:inputsdatamc}
\end{figure}

Figure~\ref{fig:inputsdatamc} shows the comparison of data and prediction for the two most discriminating input variables of the MVA, \deltarmed and \muu, used in the \njnb31 category. Good agreement is observed across all variables.

\begin{figure}[!ht]
\centering
\includegraphics[width=0.48\textwidth]{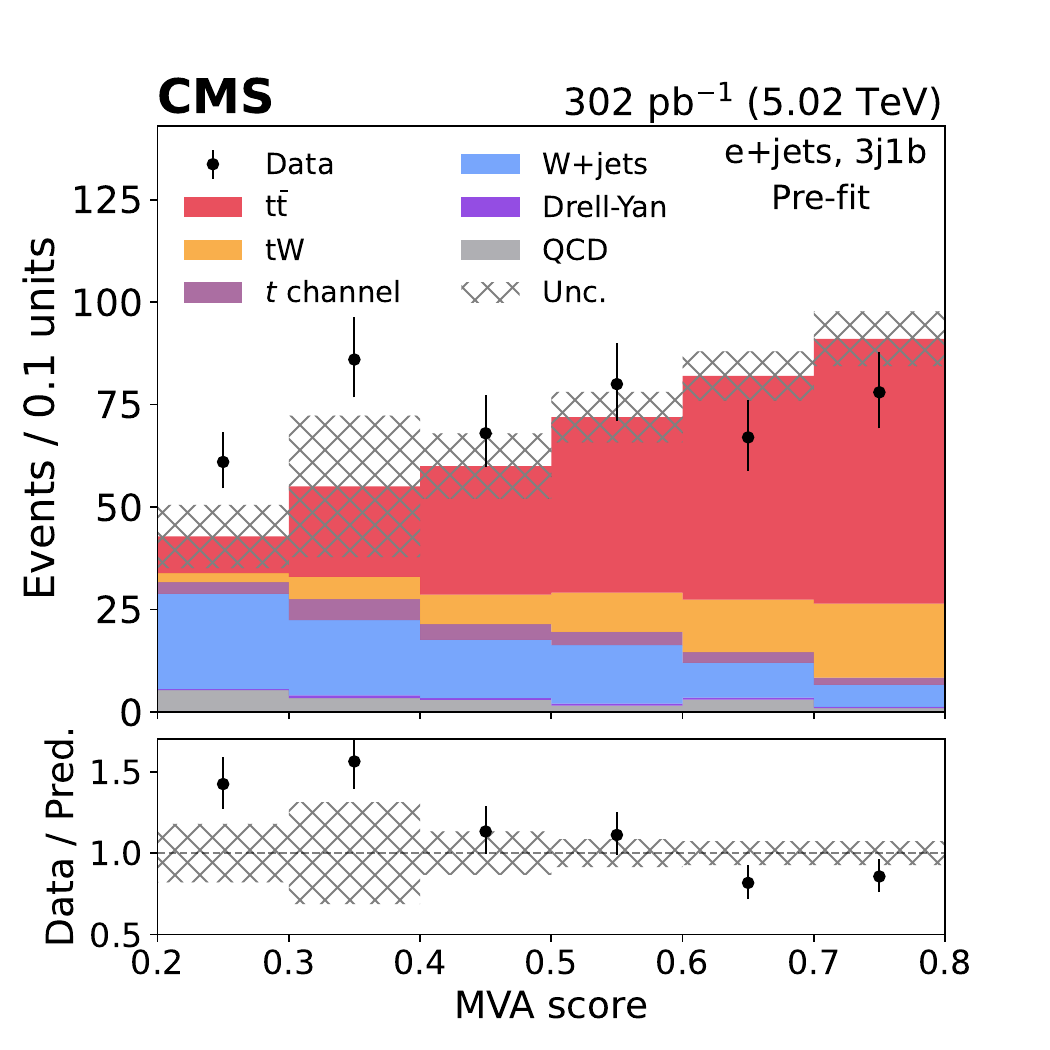}
\hfill
\includegraphics[width=0.48\textwidth]{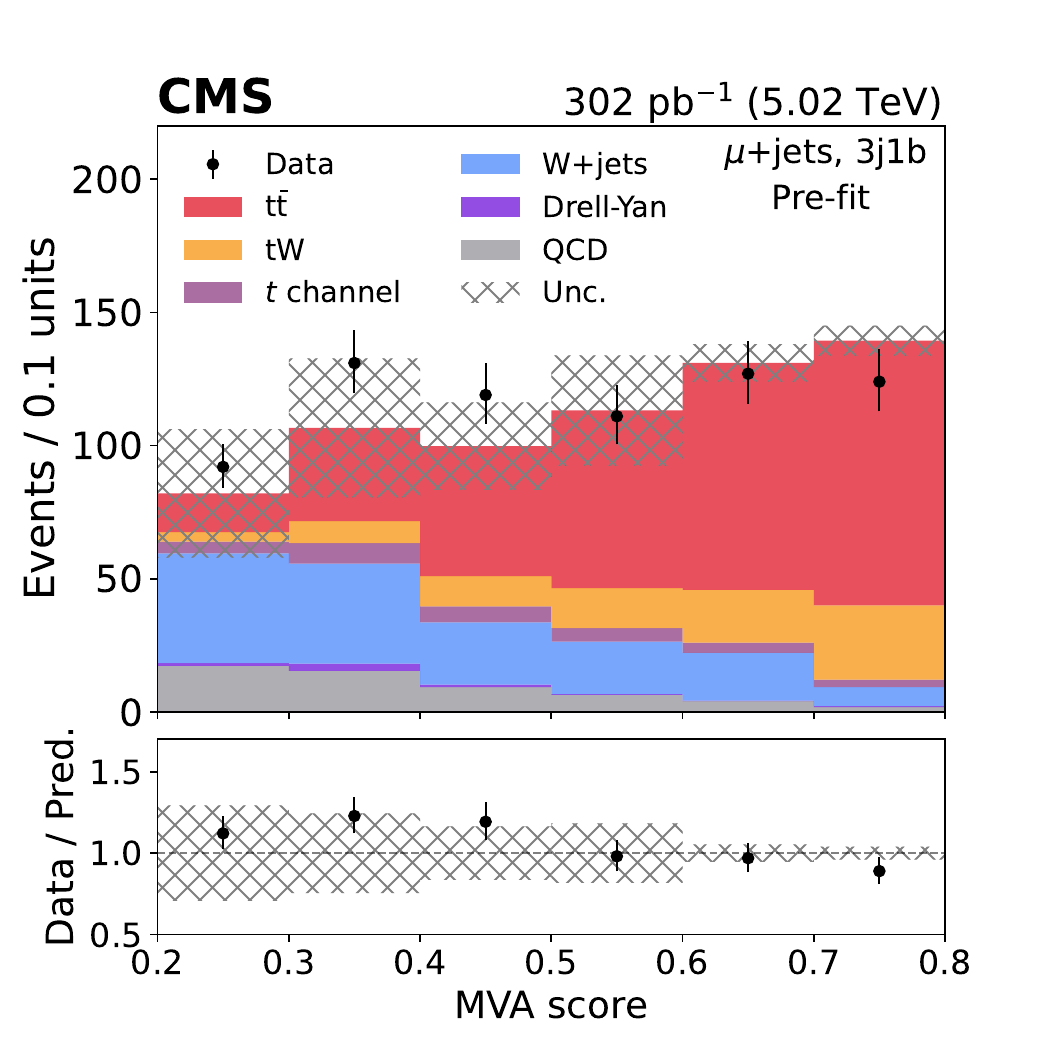}
\caption{Distributions for data and expected signal and background contributions of the MVA score for the \ejets (left) and \mjets (right) channels in the \njnb31 category, before the maximum likelihood fit. The vertical error bars represent the statistical uncertainty in the data, and the shaded band the uncertainty in the prediction. All uncertainties considered in the analysis are included in the uncertainty band. The lower panels show the data-to-prediction ratio. The first and last bins in each distribution include underflow and overflow events, respectively. The normalizations are taken with respect to the SM predictions except for QCD, which is estimated from data.}
\label{fig:mvascore}
\end{figure}

The resulting MVA score distribution is shown in Fig.~\ref{fig:mvascore} for the \ejets and \mjets final states, used then in the final fit for this category.

\section{Systematic uncertainties}
\label{sec:Systematics}

The measurement of \stt is affected by sources of systematic uncertainties related to detector effects, background normalization, running condition, and theoretical assumptions.
The methods to estimate these uncertainties are the same as those in Ref.~\cite{CMS:2021gwv}.
The experimental uncertainties (considered for all processes) mostly affect the efficiency, while modelling uncertainties (considered only for the \ttbar process) affect both the acceptance and efficiency. The modelling uncertainties account for renormalization and factorization scale choice, PS scale, ME and PS matching, PDFs, and underlying event tune. Furthermore, a normalization uncertainty is considered for each background process. The total uncertainty is calculated by combining the effects of all the individual systematic components, assuming they are independent. The sources of systematic uncertainties and the methods for their estimation are outlined below.

\begin{description}

\item{Lepton-related uncertainties:}
Lepton reconstruction and identification efficiencies are measured using \PZ boson events in simulation and low-pileup \pp collision data at 5.02\TeV~\cite{CMS:2020uim,Sirunyan:2018fpa}. Corrections are derived using the tag-and-probe method as a function of the lepton \pt and $\eta$, and are applied to the simulation to match the efficiencies in data. The uncertainty in these corrections is estimated through the variation of different parameters in the fitting templates of the method, as well as in the tag selection criteria, and is propagated to the \stt measurement as one single source per lepton flavour, including the statistical and systematic uncertainties.

The trigger efficiency is estimated using a cross-trigger technique. This consists in measuring it in data using an unbiased set of \ptmiss triggers. The related uncertainty is estimated by combining a statistical contribution, evaluated with the Clopper--Pearson binomial confidence intervals~\cite{clopper} and a systematic contribution, from MC to data efficiency ratios in several \pt, \abseta bins. It is found to be around 1--3\% for electrons and 0.8--1.6\% for muons.

\item{Jet-related uncertainties:} The contribution of the uncertainty in the jet energy scale (JES) is estimated from the change in the number of simulated events selected after changing the jet momenta by the JES uncertainties in bins of jet \pt and $\eta$, usually by no more than a few percent as described in~\cite{Khachatryan:2016kdb}. 14 different sources of uncertainties are considered, and treated as independent nuisance parameters.
The uncertainty due to the limited accuracy of the jet energy resolution (JER) is determined by changing the JER correction scale factor from its nominal value by $\pm$1 standard deviation, depending also on the \pt and $\eta$ of the jet.

\item{\ptmiss-related uncertainty:} The uncertainty in \ptmiss from the contribution of unclustered energy (energy which is not aggregated into a specific cluster in the calorimeters) is evaluated based on the momentum resolution of the different PF candidates. Also, JER, JES, and lepton energy scale effects are propagated to \ptmiss.

\item{\PQb tagging uncertainties:} In order to match the efficiency of the \PQb tagging algorithm evaluated on MC samples to that observed in data, appropriate scale factors must be applied as a function of the \pt and \abseta of the jets. Values for these factors are determined separately for heavy- and light-flavour jets, considering as heavy flavour \PQb and \PQc jets, and as light the rest~\cite{Sirunyan:2017ezt}. These scale factors are derived with an associated uncertainty, coming from different sources like JES, JER, scales. Their effect in the analysis is estimated by varying the scale factors by their uncertainties, and propagating the effect to all the categories in the analysis.

\item{L1 inefficiency:} During the data taking period corresponding to this measurement, a gradual shift in the timing of the inputs of the ECAL L1 trigger in the region $\abseta>2.0$ caused a trigger inefficiency~\cite{Sirunyan:2020zal}. Simulations are corrected to mimic this behaviour observed in data. This correction has an inherent uncertainty associated to the method used to calculate it, which is propagated to \stt by varying the correction by its associated uncertainty.

\item{Renormalization and factorization scale choice:} The uncertainty related to the missing higher-order diagrams in \POWHEG is estimated by independently varying one or both the \muf and \mur choices independently by a factor 2 or 1/2, excluding combinations with variations in opposite directions, which do not have physical meaning. As uncertainty in the signal acceptance, the maximum difference of all variations from the nominal values is assigned. 

\item{PS scale:} The effect of the choice of PS scale is studied by changing the scale used for the initial- and final-state radiation by a factor 2 and 1/2 with respect to its nominal value.

\item{ME and PS matching (\hdamp):} The effect of the ME and PS matching is taken into account using the \hdamp parameter of the \POWHEG generator, with a nominal value of $\big(1.4\,^{+0.9}_{-0.5}\big)\mt$~\cite{Sirunyan:2019dfx}. The uncertainty is estimated by varying this parameter within the uncertainties quoted above, using dedicated signal \ttbar samples. The variation with respect to the central value of the signal acceptance is taken as the uncertainty. Due to the limited size of the dedicated MC samples two separate normalization uncertainties are estimated for events with 3 or 4 jets. 

\item{Parton distribution functions:} The uncertainty arising from the choice of the proton PDF is determined by reweighing the sample of simulated signal events according to 100 replicas from the NNPDF3.1 NNLO PDF set. The eigenvectors of the Hessian matrix are taken to obtain one nuisance parameter for each replica, and the differences of the resulting variations with respect to the central value are taken as individual uncertainties, leading to 100 independent nuisance parameters. An extra contribution from the variation of \asmz corresponding to its uncertainty is considered~\cite{Butterworth:2015oua}. Since in this case there are two independent variations, two independent nuisance parameters are considered. 

\item{Underlying event tune:} The parameters of \PYTHIA related to the underlying event tune~\cite{Sirunyan:2019dfx} are adjusted to match observations in data. We consider changes in these parameters as sources of uncertainties and we use dedicated simulated signal samples accounting for these changes. The variation with respect to the central value of the signal acceptance is taken as the uncertainty. Due to the limited size of the dedicated samples, a single uncertainty is estimated for all the bins and regions of the analysis. 

\item{Background normalization:} The uncertainty in the \tW production cross section is taken to be 5.6\%, and 10\% is used for the $t$-channel single top quark production, based on the theoretical uncertainties~\cite{Campbell:2020fhf,Kidonakis:2021vob}. A normalization uncertainty of 30\% is applied to the QCD background, as discussed in Section~\ref{sec:Background}, to account for the systematic uncertainties in its estimate. Normalization uncertainties of 20 and 30\% are assigned in the \wjets and DY background estimates, respectively, following the prescription of previous analyses~\cite{Sirunyan:2017ule,wjetsref}, since the detector conditions and analysis strategy are very similar to the ones presented in this paper. In the case of \wjets, the normalization is considered as an uncorrelated parameter across the four $\Njets\Nbjets$ categories.

\item{Pileup and integrated luminosity:}  Uncertainties related to pileup are not considered since the number of pileup events is small. The uncertainty in the measurement of the integrated luminosity is estimated to be 1.9\%~\cite{bib:LUM-19-001, CMS:LUM-17-003}. This uncertainty is not included in the maximum likelihood fit, but treated as an external uncertainty and added in quadrature afterwards.
\end{description}

Most of these uncertainties, especially those coming from experimental effects, affect both the shape of the distributions and their normalization. Modelling uncertainties that are calculated with several weights, such as PDF and scale uncertainties, are estimated for each region of the analysis and applied as normalization uncertainties in each of the regions. Background normalization uncertainties and integrated luminosity uncertainty are treated as normalization uncertainties. Moreover, \hdamp and underlying event tune uncertainties, both estimated with alternative samples, are also applied as normalization uncertainties because of the limited size of the dedicated samples. In the case of \hdamp, this is done in bins of \Njets, while in the case of underlying event tune, a single normalization is used.

\section{Results}
\label{sec:Results}

\begin{figure}[!ht]
\centering
\includegraphics[width=0.75\textwidth]{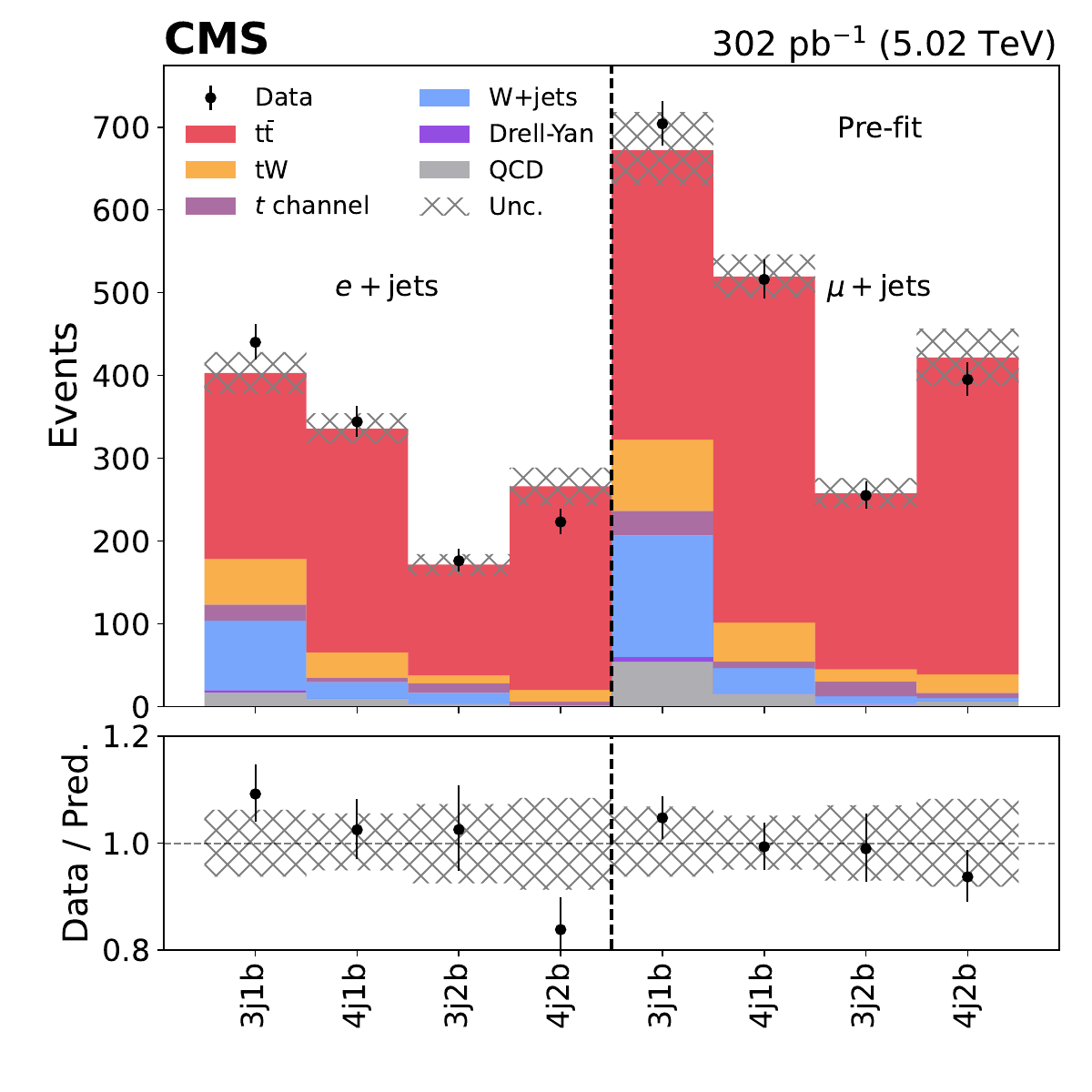}
\caption{Observed and predicted number of events in each of the eight categories of the signal region, before the maximum likelihood fit. The vertical error bars represent the statistical uncertainty in the data, and the shaded band the  uncertainty in the prediction. All uncertainties considered in the analysis are included in the uncertainty band. The lower panels show the data-to-prediction ratio. The normalizations are taken with respect to the SM predictions except for QCD, which is estimated from data.}
\label{fig:master}
\end{figure}

After the event selection and categorization described above, the resulting observed number of events in data and the expected number of events for signal and background in each
of the eight categories in the signal region are those shown in Fig.~\ref{fig:master}.

\begin{figure}[!p]
\centering
\includegraphics[width=0.9\textwidth]{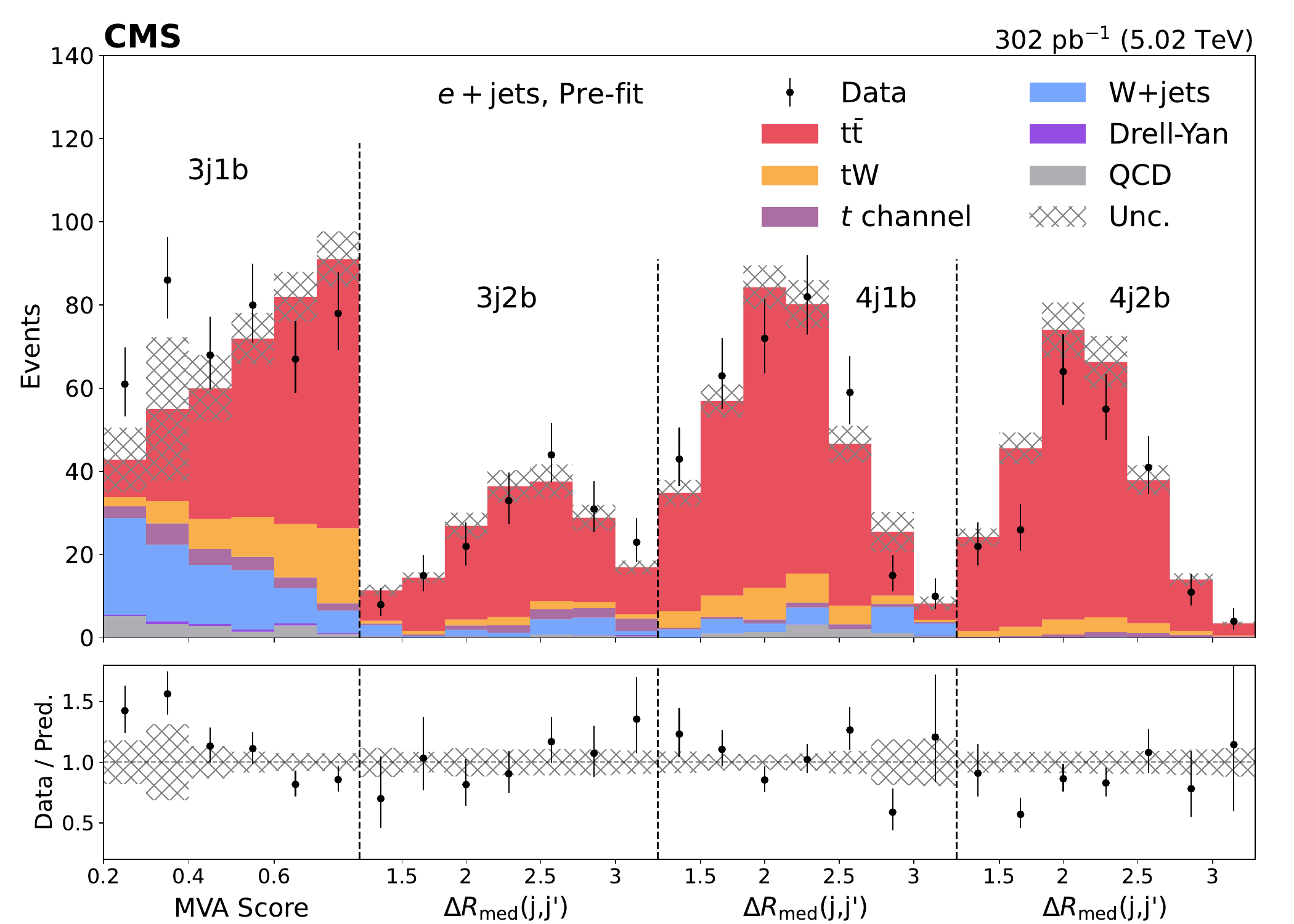} \\
\includegraphics[width=0.9\textwidth]{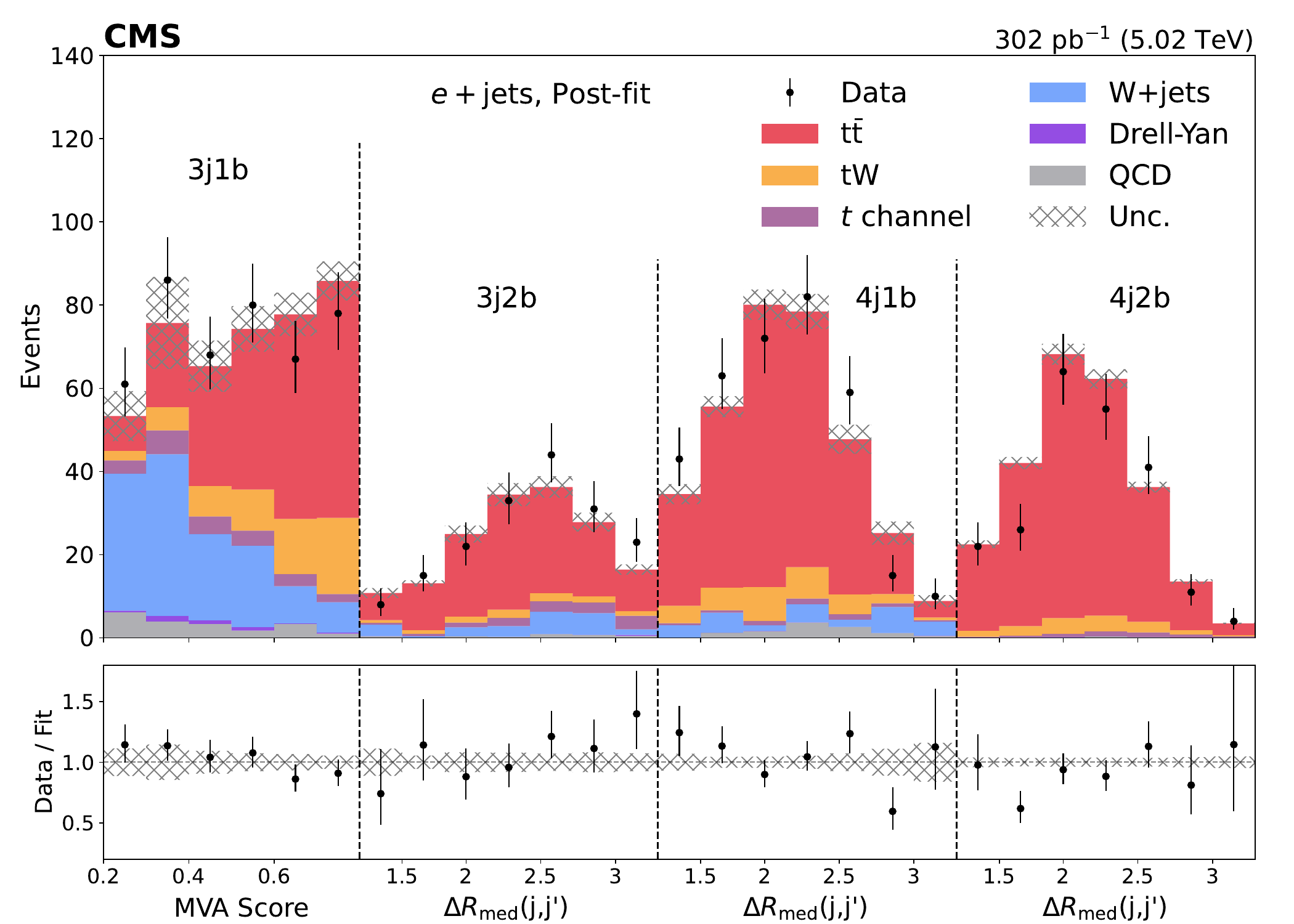}
\caption{Distributions for the \ejets final state before (upper plot) and after (lower plot) the maximum likelihood fit: MVA score bins for the \njnb31 category and \deltarmed bins for the other categories. The vertical error bars represent the statistical uncertainty in the data, and the shaded band the uncertainty of the prediction. All uncertainties considered in the analysis are included in the uncertainty band. The lower panels show the ratio of data to prediction or data to fit, respectively. The first and last bins in each distribution include underflow and overflow events, respectively.}
\label{fig:prefite}
\end{figure}

\begin{figure}[!p]
\centering
\includegraphics[width=0.9\textwidth]{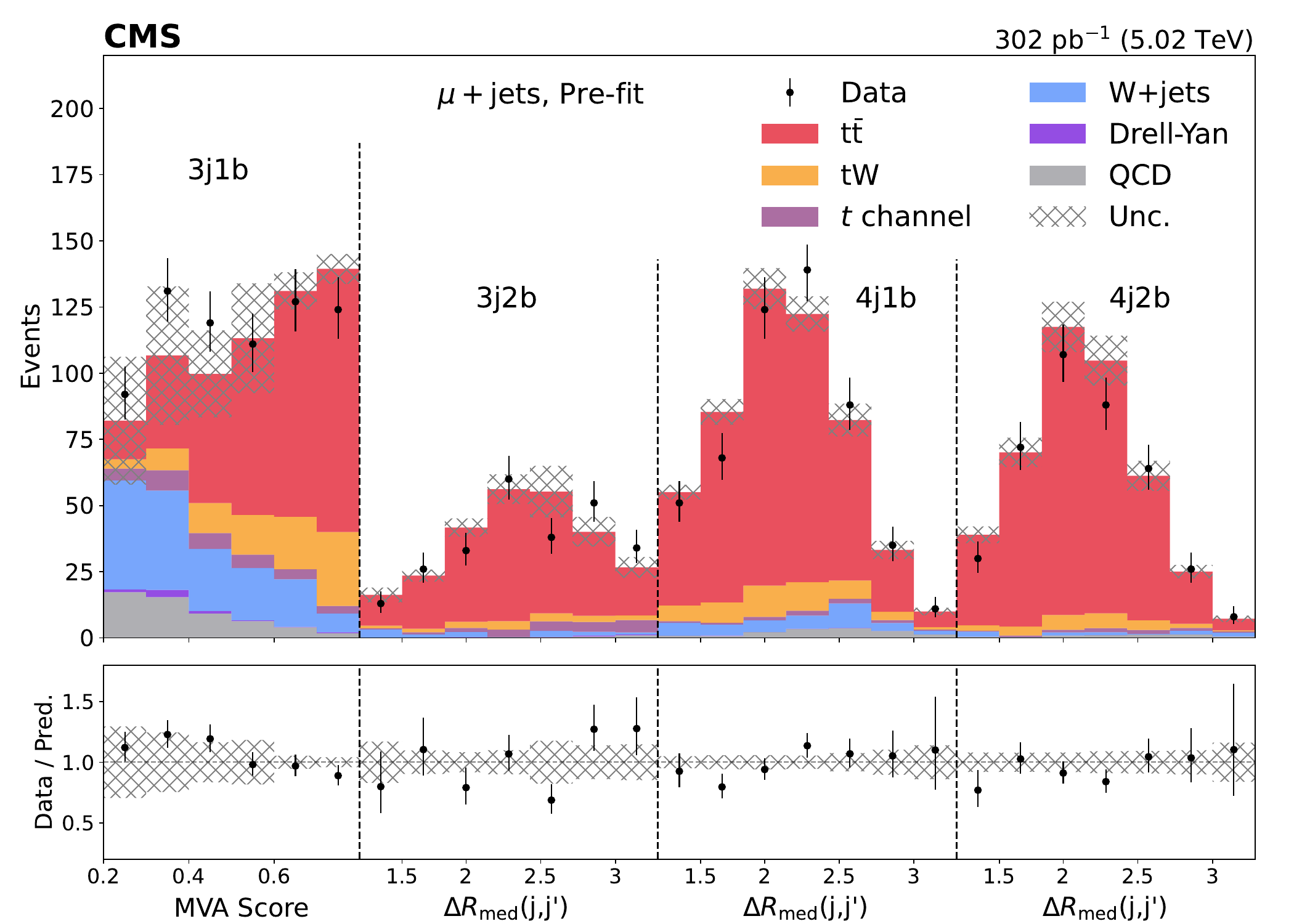} \\
\includegraphics[width=0.9\textwidth]{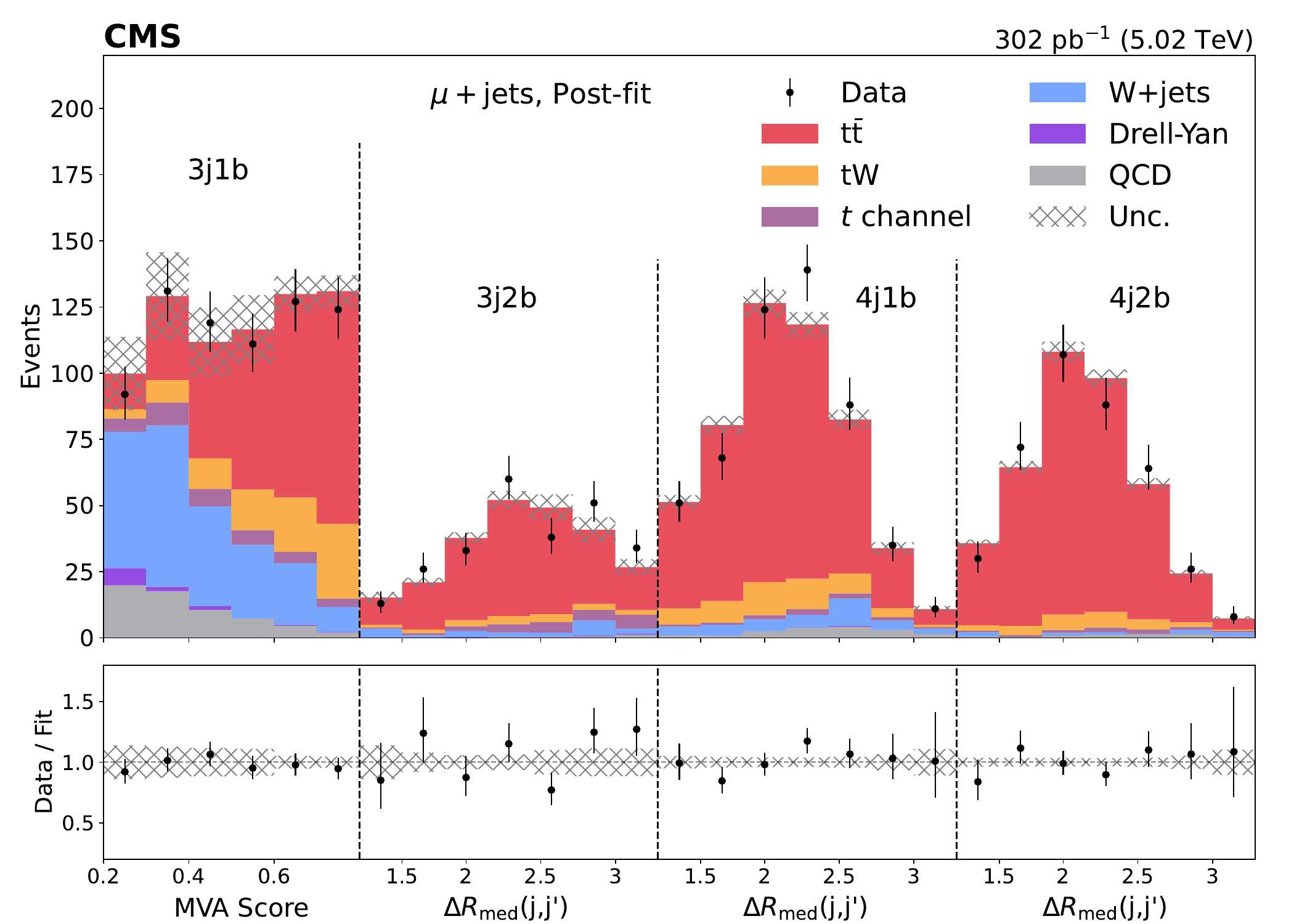}
\caption{Distributions for the \mjets final state before (upper plot) and after (lower plot) the maximum likelihood fit: MVA score bins for the \njnb31 category and \deltarmed bins for the other categories. The vertical error bars represent the statistical uncertainty in the data, and the shaded band the uncertainty of the prediction. All uncertainties considered in the analysis are included in the uncertainty band. The lower panels show the ratio of data to prediction or data to fit, respectively. The first and last bins in each distribution include underflow and overflow events, respectively.}
\label{fig:prefitmu}
\end{figure}

{\tolerance=1200
The cross section is obtained by performing a profiled maximum likelihood fit to the \deltarmed distribution in the different jet/\PQb-jet multiplicity and lepton flavour categories, except for the \njnb31 category, where the MVA score distribution is used. The cross section value is fitted simultaneously with the nuisance parameters related to the systematic uncertainties. The choice of the \deltarmed distribution is motivated by the fact that this observable is particularly sensitive to the kinematics of \ttbar production: in fact, the signal peaks more sharply at intermediate \deltarmed than the background. Using the MVA score distribution instead of \deltarmed for the \njnb31 category results in a relative reduction of 13\% in the final uncertainty of the result. In the fit, seven bins are used for the \deltarmed and six bins for the MVA score distributions, for a total of 54 bins (27 per lepton flavour). The most sensitive category is found to be the {\PGm}+\njnb31 one. The signal strength ($r$), \ie the ratio of the observed to the predicted cross section is extracted and uncertainties are constrained. The statistical uncertainty coming from the limited size of the MC samples is incorporated into the maximum likelihood fit using the Barlow--Beeston ``light'' method~\cite{Conway:2011in}. The MVA score and \deltarmed distributions for the \ejets and \mjets final states before and after the fit are shown in Figs.~\ref{fig:prefite} and~\ref{fig:prefitmu}.
\par}

The signal strength is extracted from a profile-likelihood fit using the CMS \textsc{combine} tool~\cite{cmscollaboration2024cms} and a cross section of $\stt=62.5\pb$ is measured.
The systematic uncertainty amounts to ${}^{+4.2\%}_{-4.0\%}$, the statistical uncertainty to $\pm$2.6\%, and the integrated luminosity uncertainty to $\pm$1.9\%. The measurement is also done in \ejets ($\stt=61.8\pm4.6\pb$) and \mjets ($\stt=63.6\pm3.6\pb$) separately, showing consistent cross section with the combined fit.

\begin{figure}[!ht]
\centering
\includegraphics[width=0.8\textwidth]{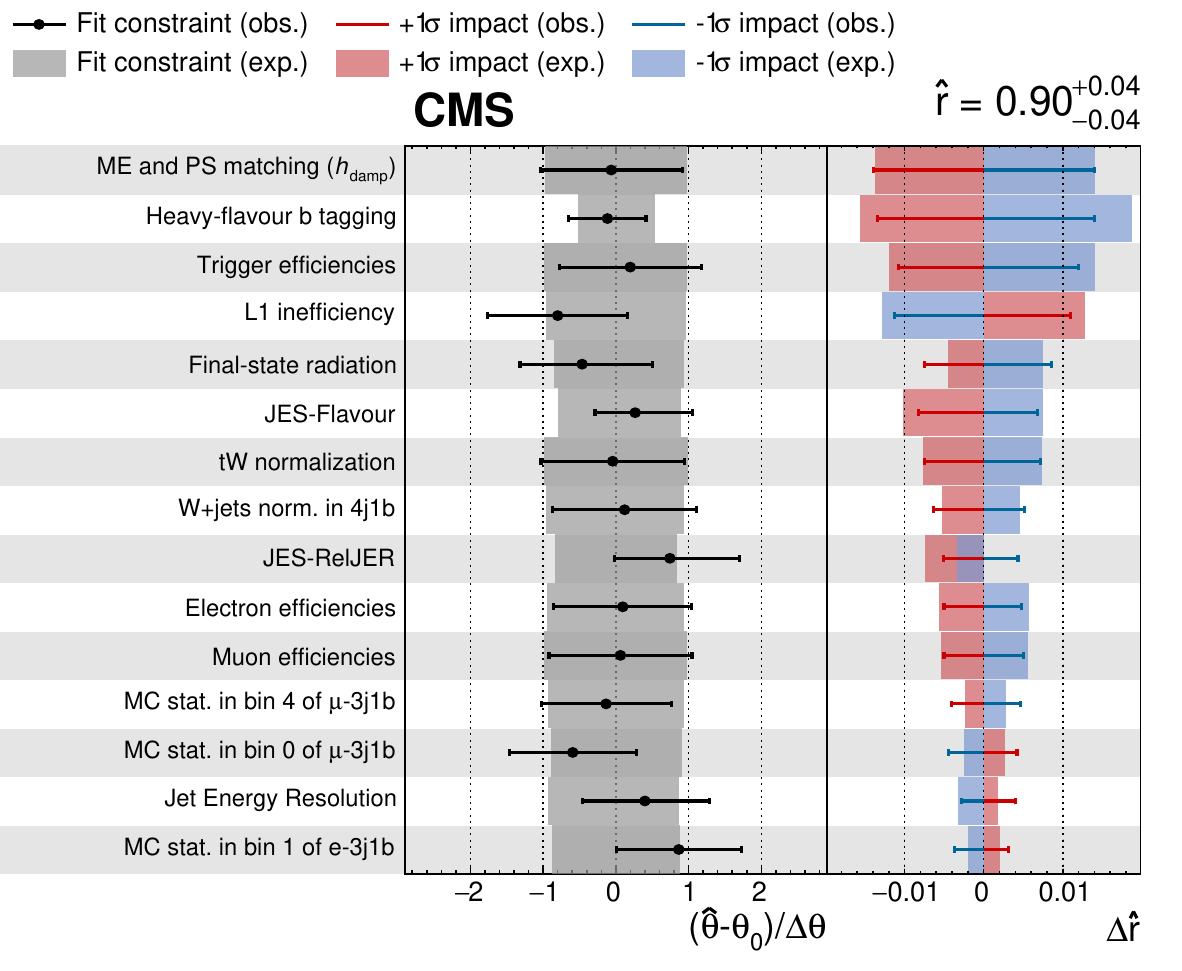}
\caption{The largest impacts on the signal strength, $\Delta\hat{r}$ (right column) and ratios $(\hat{\theta}-\theta_0)/\Delta\theta$ (middle column) for the nuisance parameters listed in the left column from the maximum likelihood fit used to determine the \ttbar cross section. The horizontal bars in the rightmost plot show the ratio of the uncertainties of the fit result to the pre-fit ones, effectively giving the constraint on the nuisance parameter. The JES uncertainties are divided into several sources, where ''JES-Flavour'' comes from the corrections applied to correct the different detector response to gluon and quark jets, and
''JES-RelJER'' accounts for the $\eta$ dependence uncertainty from jet \pt resolution.}
\label{fig:impacts1l}
\end{figure}

Figure~\ref{fig:impacts1l} shows the impact of the different systematic uncertainties on the signal strength and the constraints on the nuisance parameters. The impact is defined as the shift $\Delta\hat{r}$ induced in $r$ when the nuisance parameter $\theta$ is varied by $\pm$1 standard deviation ($\sigma$) around the value obtained for it after the fit. The leading uncertainties are those on the heavy-flavour \PQb tagging efficiency, the ME and PS matching, and the trigger efficiency. Figure~\ref{fig:impacts1l} also shows the ratios $(\hat{\theta}-\theta_0)/\Delta\theta$, where $\hat{\theta}$ and $\theta_0$ are the values of a given nuisance parameter after and before the fit, and $\Delta\theta$ the corresponding uncertainty before the fit. The uncertainties coming from the ME and PS matching, and heavy-flavour \PQb tagging efficiencies appear as the dominant ones.

Alternative \ttbar MC samples generated with $\mt=166.5$ and 178.5\GeV are used to estimate, by linear interpolation, the variation in the \ttbar cross section when \mt changes.
The measured cross section varies by $\mp$0.11\pb when the top quark mass is varied from its nominal value of 172.5\GeV within its uncertainty of $\pm$0.33\GeV~\cite{TOP-22-001}.

The result in the lepton+jets channel is combined with the result in the dilepton channel from Ref.~\cite{CMS:2021gwv}.
The sources of systematic uncertainty are considered as fully correlated, with the exception of few uncertainties (JES, PDF, and \asmz) that are treated as uncorrelated. Namely, in the dilepton channel a single source per uncertainty is considered whereas in the lepton+jets, 14, 100, and 2 sources per uncertainty are considered, respectively. Moreover, there are other uncertainties which were not considered
in the dilepton analysis, such as \PQb tagging, unclustered energy, and QCD and $t$-channel single top quark production normalizations.
As in the lepton+jets case, a profile likelihood fit is performed on the data and a cross section of $\stt=62.3\pb$ is measured. The systematic uncertainty amounts to $\pm$3.9\%, the statistical uncertainty to $\pm$2.4\%, and the integrated luminosity uncertainty to $\pm$1.9\%. As expected, this result is dominated by the lepton+jets measurement, given its significantly smaller uncertainty.

\begin{figure}[!ht]
\centering
\includegraphics[width=0.9\textwidth]{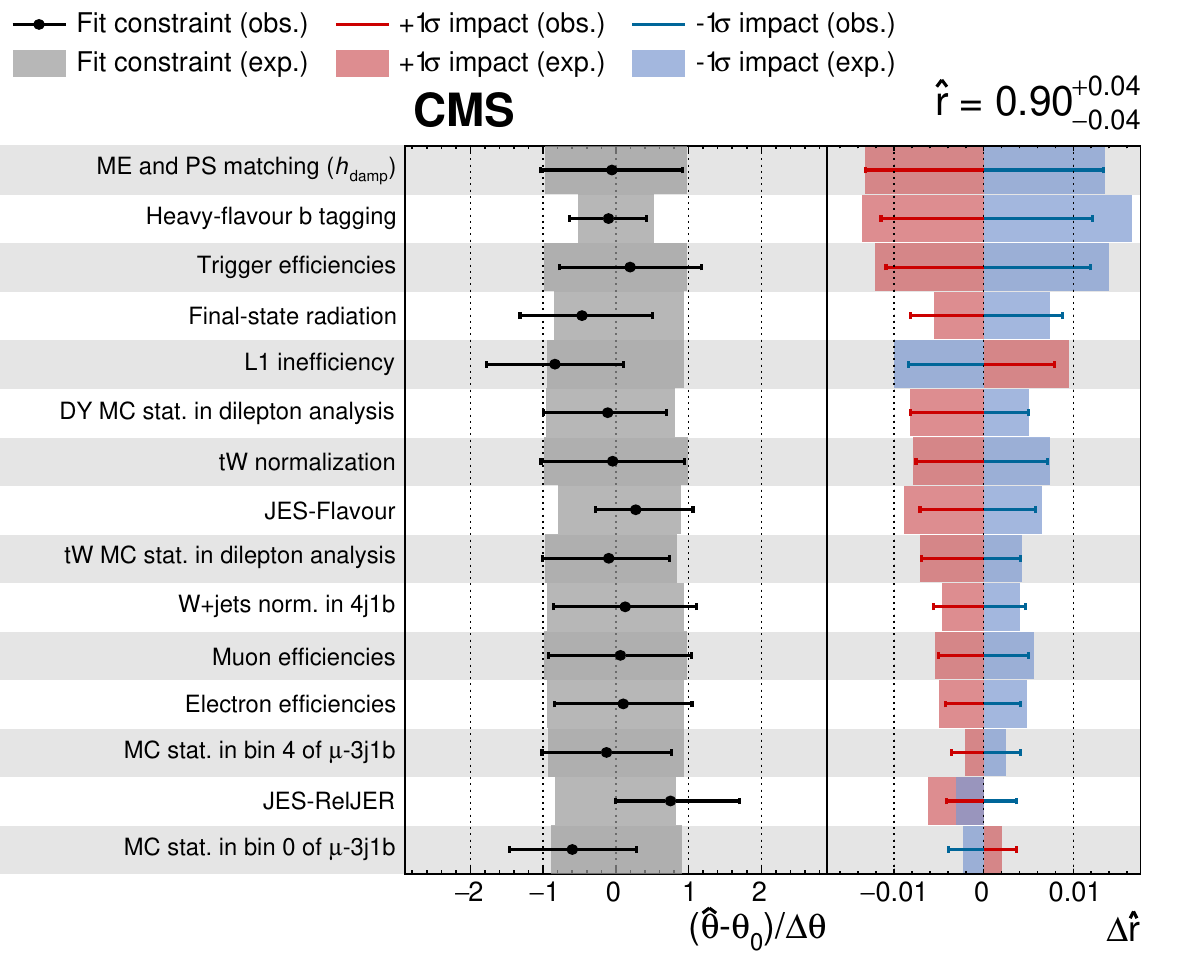}
\caption{The largest impacts on the signal strength, $\Delta\hat{r}$ (right column) and ratios $(\hat{\theta}-\theta_{0})/\Delta\theta$ (middle column) for the nuisance parameters listed in the left column from the maximum likelihood fit in the combination with the dilepton result, used to determine the \ttbar cross section. The horizontal bars in the rightmost plot show the ratio of the uncertainties of the fit result to the pre-fit ones, effectively giving the constraint on the nuisance parameter. The JES uncertainties are divided into several sources, accounting for different effects, where ''JES-Flavour'' comes from the
corrections applied to correct the different detector response to gluon and quark jets, and ''JES-RelJER'' accounts for the $\eta$ dependence uncertainty from jet \pt resolution.}
\label{fig:impactscomb}
\end{figure}

Figure~\ref{fig:impactscomb} shows the impact of the systematic uncertainties on the signal strength and the ratios $(\hat{\theta}-\theta_0)/\Delta\theta$ for the nuisance parameters as determined by the combined fit. In this case, the ME and PS matching, the heavy-flavour \PQb tagging efficiency and the trigger efficiencies are still among the leading uncertainties. As in the lepton+jets analysis, the uncertainty coming from the heavy-flavour \PQb tagging efficiency appears as one of the most constrained ones.

\begin{figure}[!ht]
\centering
\includegraphics[width=0.9\textwidth]{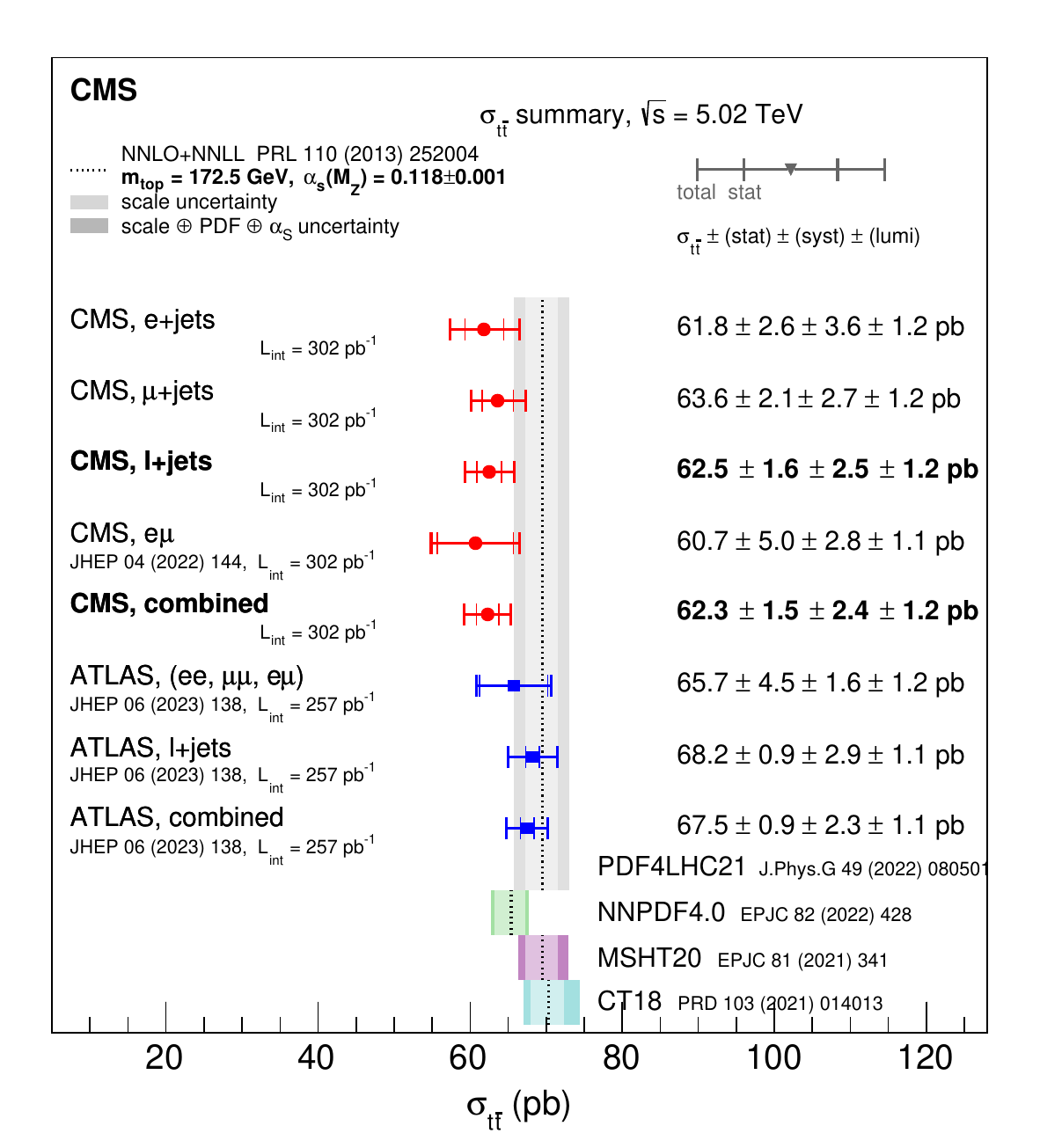}
\caption{Summary of the most recent measurements from the ATLAS and CMS Collaborations using data collected at $\sqrt{s}=5.02\TeV$. In the plot also several theoretical predictions are shown: the current prediction~\cite{theor1} (calculated with the PDF4LHC21 set) and other previous predictions using different PDF sets~\cite{theor2,theor3,theor4}.}
\label{fig:summary}
\end{figure}

The results presented in this analysis are consistent within two standard deviations with the SM prediction. These results, together with previous CMS and ATLAS measurements, are shown in Fig.~\ref{fig:summary}.

\section{Summary}
\label{sec:Summary}

A measurement of the top quark pair production cross section in proton-proton collisions at a centre-of-mass energy of 5.02\TeV is performed
for events with one electron or one muon and multiple jets using data collected by the CMS experiment in 2017 in conditions of low number of additional interactions per bunch crossing, corresponding to an integrated luminosity of 302\pbinv. The dominant background sources in the analysis are \wjets and tW processes, whose contribution is estimated from simulation. In addition, the contribution from quantum chromodynamics multijet events is estimated from data. The cross section is measured using a maximum likelihood fit to eight event categories defined in terms of the number of jets, \PQb-tagged jets and lepton flavour.
The cross section is found to be $62.5\pm1.6\stat\,^{+2.6}_{-2.5}\syst\pm1.2\lum\pb$.
This measurement is combined with the result obtained in the dilepton channel, based on the same data set, resulting in a value
of $62.3\pm1.5\stat\pm2.4\syst\pm1.2\lum\pb$. In both cases, the dominant uncertainties are those associated with the integrated luminosity and with the \PQb tagging scale factors for heavy flavours.
These values are consistent within two standard deviations with the standard model prediction and of previous measurements from CMS and ATLAS.

\begin{acknowledgments}
We congratulate our colleagues in the CERN accelerator departments for the excellent performance of the LHC and thank the technical and administrative staffs at CERN and at other CMS institutes for their contributions to the success of the CMS effort. In addition, we gratefully acknowledge the computing centres and personnel of the Worldwide LHC Computing Grid and other centres for delivering so effectively the computing infrastructure essential to our analyses. Finally, we acknowledge the enduring support for the construction and operation of the LHC, the CMS detector, and the supporting computing infrastructure provided by the following funding agencies: SC (Armenia), BMBWF and FWF (Austria); FNRS and FWO (Belgium); CNPq, CAPES, FAPERJ, FAPERGS, and FAPESP (Brazil); MES and BNSF (Bulgaria); CERN; CAS, MoST, and NSFC (China); MINCIENCIAS (Colombia); MSES and CSF (Croatia); RIF (Cyprus); SENESCYT (Ecuador); ERC PRG, RVTT3 and MoER TK202 (Estonia); Academy of Finland, MEC, and HIP (Finland); CEA and CNRS/IN2P3 (France); SRNSF (Georgia); BMBF, DFG, and HGF (Germany); GSRI (Greece); NKFIH (Hungary); DAE and DST (India); IPM (Iran); SFI (Ireland); INFN (Italy); MSIP and NRF (Republic of Korea); MES (Latvia); LMTLT (Lithuania); MOE and UM (Malaysia); BUAP, CINVESTAV, CONACYT, LNS, SEP, and UASLP-FAI (Mexico); MOS (Montenegro); MBIE (New Zealand); PAEC (Pakistan); MES and NSC (Poland); FCT (Portugal); MESTD (Serbia); MCIN/AEI and PCTI (Spain); MOSTR (Sri Lanka); Swiss Funding Agencies (Switzerland); MST (Taipei); MHESI and NSTDA (Thailand); TUBITAK and TENMAK (Turkey); NASU (Ukraine); STFC (United Kingdom); DOE and NSF (USA).

\hyphenation{Rachada-pisek} Individuals have received support from the Marie-Curie programme and the European Research Council and Horizon 2020 Grant, contract Nos.\ 675440, 724704, 752730, 758316, 765710, 824093, 101115353, 101002207, and COST Action CA16108 (European Union); the Leventis Foundation; the Alfred P.\ Sloan Foundation; the Alexander von Humboldt Foundation; the Science Committee, project no. 22rl-037 (Armenia); the Belgian Federal Science Policy Office; the Fonds pour la Formation \`a la Recherche dans l'Industrie et dans l'Agriculture (FRIA-Belgium); the F.R.S.-FNRS and FWO (Belgium) under the ``Excellence of Science -- EOS" -- be.h project n.\ 30820817; the Beijing Municipal Science \& Technology Commission, No. Z191100007219010 and Fundamental Research Funds for the Central Universities (China); the Ministry of Education, Youth and Sports (MEYS) of the Czech Republic; the Shota Rustaveli National Science Foundation, grant FR-22-985 (Georgia); the Deutsche Forschungsgemeinschaft (DFG), among others, under Germany's Excellence Strategy -- EXC 2121 ``Quantum Universe" -- 390833306, and under project number 400140256 - GRK2497; the Hellenic Foundation for Research and Innovation (HFRI), Project Number 2288 (Greece); the Hungarian Academy of Sciences, the New National Excellence Program - \'UNKP, the NKFIH research grants K 131991, K 133046, K 138136, K 143460, K 143477, K 146913, K 146914, K 147048, 2020-2.2.1-ED-2021-00181, and TKP2021-NKTA-64 (Hungary); the Council of Science and Industrial Research, India; ICSC -- National Research Centre for High Performance Computing, Big Data and Quantum Computing and FAIR -- Future Artificial Intelligence Research, funded by the NextGenerationEU program (Italy); the Latvian Council of Science; the Ministry of Education and Science, project no. 2022/WK/14, and the National Science Center, contracts Opus 2021/41/B/ST2/01369 and 2021/43/B/ST2/01552 (Poland); the Funda\c{c}\~ao para a Ci\^encia e a Tecnologia, grant CEECIND/01334/2018 (Portugal); the National Priorities Research Program by Qatar National Research Fund; MCIN/AEI/10.13039/501100011033, ERDF ``a way of making Europe", and the Programa Estatal de Fomento de la Investigaci{\'o}n Cient{\'i}fica y T{\'e}cnica de Excelencia Mar\'{\i}a de Maeztu, grant MDM-2017-0765 and Programa Severo Ochoa del Principado de Asturias (Spain); the Chulalongkorn Academic into Its 2nd Century Project Advancement Project, and the National Science, Research and Innovation Fund via the Program Management Unit for Human Resources \& Institutional Development, Research and Innovation, grant B39G670016 (Thailand); the Kavli Foundation; the Nvidia Corporation; the SuperMicro Corporation; the Welch Foundation, contract C-1845; and the Weston Havens Foundation (USA).
\end{acknowledgments}

\bibliography{auto_generated}
\cleardoublepage \appendix\section{The CMS Collaboration \label{app:collab}}\begin{sloppypar}\hyphenpenalty=5000\widowpenalty=500\clubpenalty=5000
\cmsinstitute{Yerevan Physics Institute, Yerevan, Armenia}
{\tolerance=6000
A.~Hayrapetyan, A.~Tumasyan\cmsAuthorMark{1}\cmsorcid{0009-0000-0684-6742}
\par}
\cmsinstitute{Institut f\"{u}r Hochenergiephysik, Vienna, Austria}
{\tolerance=6000
W.~Adam\cmsorcid{0000-0001-9099-4341}, J.W.~Andrejkovic, T.~Bergauer\cmsorcid{0000-0002-5786-0293}, S.~Chatterjee\cmsorcid{0000-0003-2660-0349}, K.~Damanakis\cmsorcid{0000-0001-5389-2872}, M.~Dragicevic\cmsorcid{0000-0003-1967-6783}, P.S.~Hussain\cmsorcid{0000-0002-4825-5278}, M.~Jeitler\cmsAuthorMark{2}\cmsorcid{0000-0002-5141-9560}, N.~Krammer\cmsorcid{0000-0002-0548-0985}, A.~Li\cmsorcid{0000-0002-4547-116X}, D.~Liko\cmsorcid{0000-0002-3380-473X}, I.~Mikulec\cmsorcid{0000-0003-0385-2746}, J.~Schieck\cmsAuthorMark{2}\cmsorcid{0000-0002-1058-8093}, R.~Sch\"{o}fbeck\cmsorcid{0000-0002-2332-8784}, D.~Schwarz\cmsorcid{0000-0002-3821-7331}, M.~Sonawane\cmsorcid{0000-0003-0510-7010}, W.~Waltenberger\cmsorcid{0000-0002-6215-7228}, C.-E.~Wulz\cmsAuthorMark{2}\cmsorcid{0000-0001-9226-5812}
% ...
\par}
\vskip\cmsinstskip
\dag:~Deceased\\
$^{1}$Also at Yerevan State University, Yerevan, Armenia\\
$^{2}$Also at TU Wien, Vienna, Austria\\
% ...
\end{sloppypar}
\end{document}